\documentclass[slac_one]{revtex4}
\usepackage{graphicx}
\usepackage{fancyhdr}
\begin{document}

\title{Recent results on angular momentum and strangeness in the nucleon} 

%

\author{A. Hillenbrand (on behalf of the HERMES Collaboration)}
\affiliation{DESY, 15738 Zeuthen, Germany}
%

\begin{abstract}
HERMES has measured azimuthal single-spin asymmetries of pions and charged kaons
produced in deep-inelastic scattering of electrons and positrons off a transversely
polarized hydrogen target.
These asymmetries provide information about the Collins and Sivers mechanisms,
which are signals for the transverse parton distribution function 
$h_1^q$ in convolution with the Collins fragmentation function $H_1^{\perp,q}$, and of the
Sivers distribution function $f_{1T}^{\perp,q}$ with the spin-averaged fragmentation function $D_q^h$, respectively.
Furthermore, both the momentum and helicity distributions of the strange quark sea have been extracted in leading order from the multiplicity and the inclusive double spin asymmetry, respectively, in the production of charged kaons when scattering polarized positrons off a longitudinally polarized deuterium target.
The shape of the momentum distribution is softer than that of the average of the $\bar{u}$ and $\bar{d}$ quarks. 
The helicity distribution is found to be consistent with zero.
\end{abstract}

\maketitle

\thispagestyle{fancy}


\section{SIVERS AND COLLINS MOMENTS} 
\label{Sec:Trans}

In leading twist, three parton distribution functions (PDFs) survive the integration over intrinsic transverse momenta: the momentum distribution $q(x)$, the helicity distribution $\Delta q(x)$ \cite{hermes_dq} and the transversity distribution $\delta q(x)\equiv h_1^q(x)$.
The latter two describe the correlation between quark and nucleon spin for longitudinally and transversely polarized nucleons, respectively.
In the helicity basis, the transversity distribution $h_1^q(x)$ is related to a forward quark-nucleon scattering amplitude involving helicity flips of both nucleon and quark. As such, transversity is chiral-odd and can not be measured in inclusive deep-inelastic scattering (DIS).
However, it is possible to measure $h_1^q(x)$ in semi-inclusive DIS (SIDIS) in combination with another chiral-odd object, the Collins fragmentation function, which relates the transverse polarization of the struck quark with the transverse momentum $P_{h \perp}$ of the produced hadron. This so-called {\it Collins mechanism} \cite{Collins} manifests itself in a single-spin asymmetry (SSA), i.e. a left-right asymmetry in the production of hadrons in the plane transverse to the direction of the virtual photon. 

Another SSA observable in SIDIS on a transversely polarized target originates from the {\it Sivers mechanism} \cite{Sivers}: it emerges from the combination of the spin-averaged fragmentation function with the Sivers distribution function.
This naive time-reversal odd function describes the correlation between the transverse polarization of the nucleon with the transverse momentum $p_{\perp}$ of its quarks.
A non-zero Sivers function implies non-vanishing quark orbital angular momentum in the nucleon wave function.

Both mechanisms were studied at HERMES using a transversely polarized hydrogen gas target internal to the HERA electron/positron storage ring that was operated with a beam energy of $E=27.6\,\mathrm{GeV}$.
The presented results include the full data set recorded with transverse target polarization in the years 2002-2005 \cite{Markus}.
Semi-inclusive DIS events were selected by requiring the squared invariant mass of the virtual photon $Q^2>1\,\mathrm{GeV}^2$, the fractional energy transfer to the target $y=\nu/E<0.95$ (with $\nu$ being the virtual photon energy), the squared invariant mass in the final state $W^2>10\,\mathrm{GeV}^2$ and $0.2<z<0.7$, where $z=E_h/\nu$ with $E_h$ the energy of the produced hadron in the target rest frame.
In addition, hadrons were constrained to the momentum range $2\, \mathrm{GeV}<P<15\,\mathrm{GeV}$, where they could be identified as $\pi^{\pm}$, $K^{\pm}$ and (anti-)protons by the
dual-radiator ring imaging \v{C}erenkov (RICH) detector.
The azimuthal single-spin asymmetries
\begin{equation}
	A_{\mathrm{UT}}(\phi, \phi_S, \dots)  =  \frac{1}{\langle P_z \rangle} 
	\frac{\sigma^{\Uparrow}-\sigma^{\Downarrow}}{\sigma^{\Uparrow}+\sigma^{\Downarrow}}
\end{equation}
(where '$\Uparrow$'/'$\Downarrow$' denotes the target spin vector pointing up/down)
 have been extracted in terms of the angles $\phi$ and $\phi_S$, which are defined as the angle between the lepton scattering plane and either the hadron production plane ($\phi$) or the target spin direction ($\phi_S$), respectively.
The Collins amplitude $\langle \sin\left( \phi + \phi_s \right) \rangle^h_{\mathrm{UT}}$ and Sivers amplitude $\langle \sin\left( \phi - \phi_s \right) \rangle^h_{\mathrm{UT}}$ have distinctive angular signatures and were extracted simultaneously using maximum likelihood fits.
The fits also allowed for other possible sine modulations coming from the spin-dependent cross section.

\begin{figure*}[t]
\centering
\includegraphics[width=0.49\textwidth]{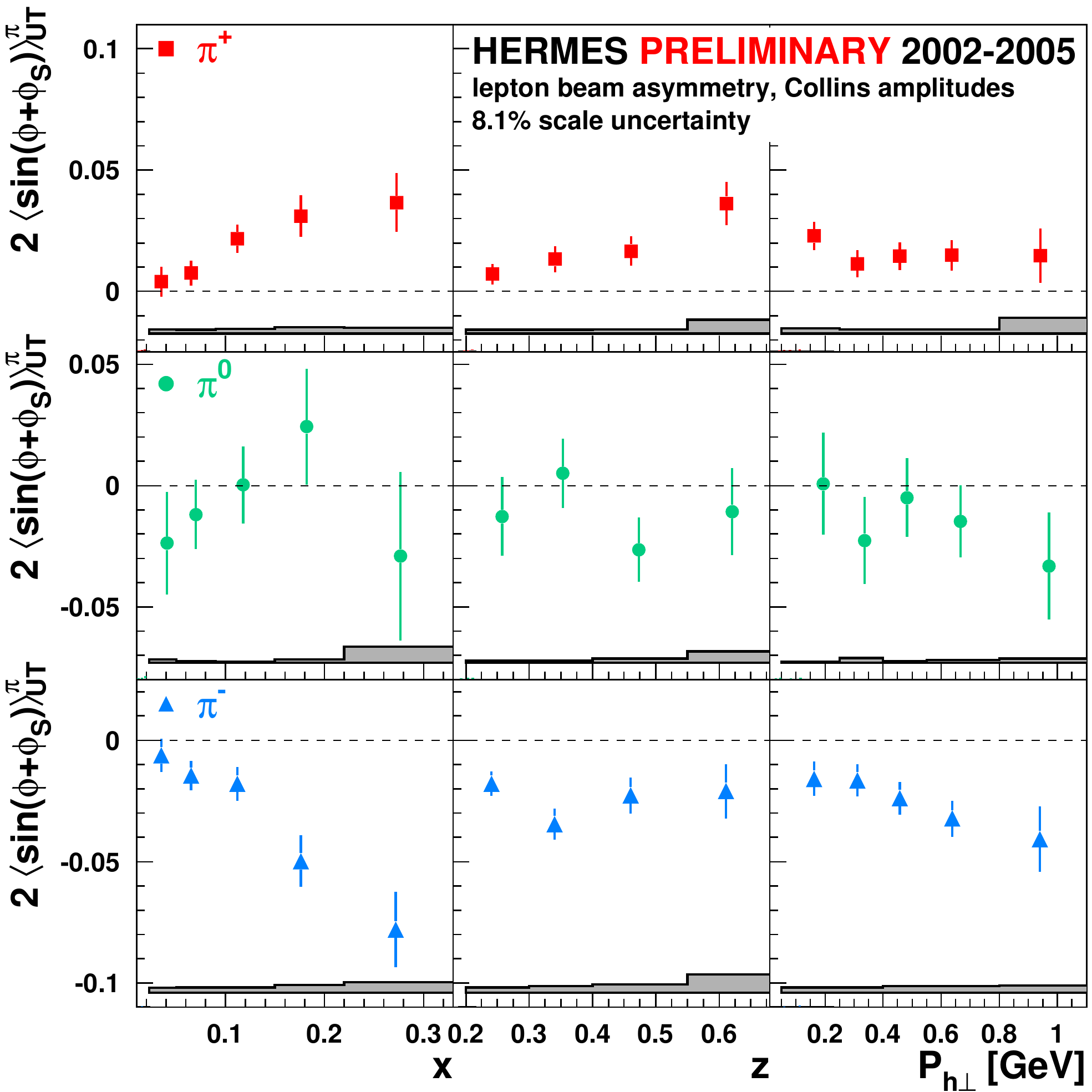}
\includegraphics[width=0.49\textwidth]{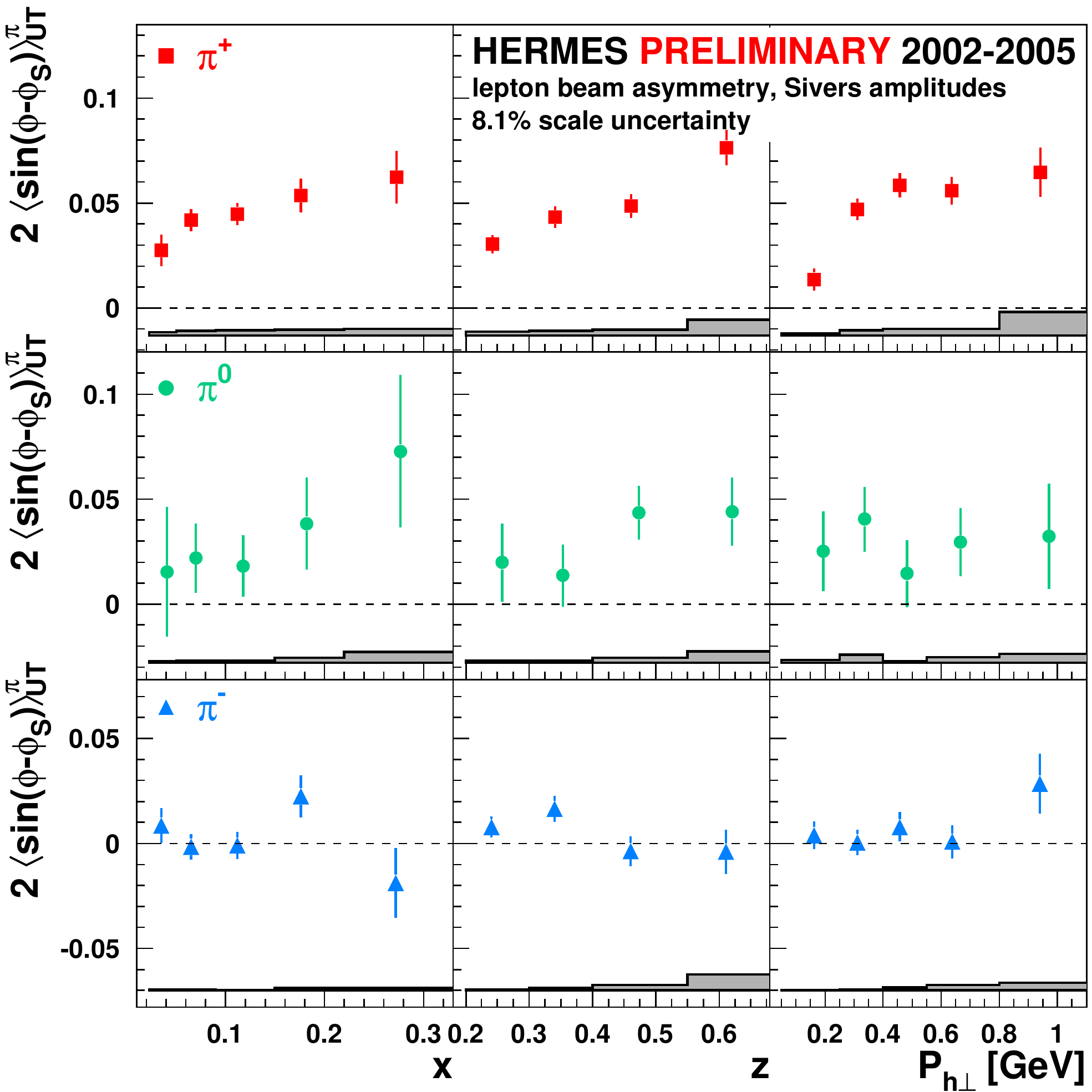}
\caption{Collins (left) and Sivers (right) amplitudes for pions as function of $x$, $z$ and $P_{h\perp}$. The error bands give the maximal systematic error; in addition there is a common 8.1\% scaling uncertainty due to the target polarization uncertainty.} \label{Fig:Pions}
\end{figure*}

\begin{figure*}[t]
\centering
\includegraphics[width=0.49\textwidth]{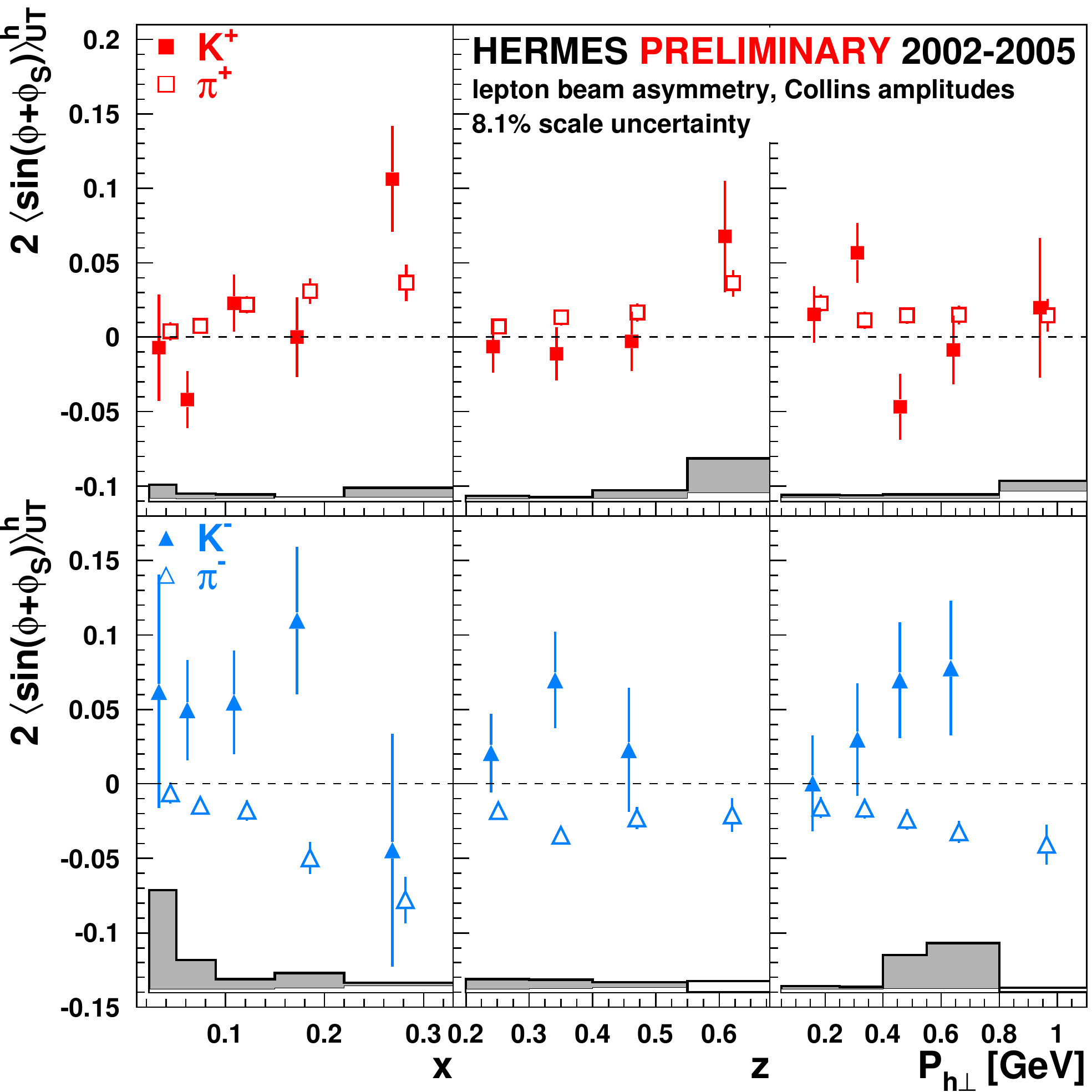}
\includegraphics[width=0.49\textwidth]{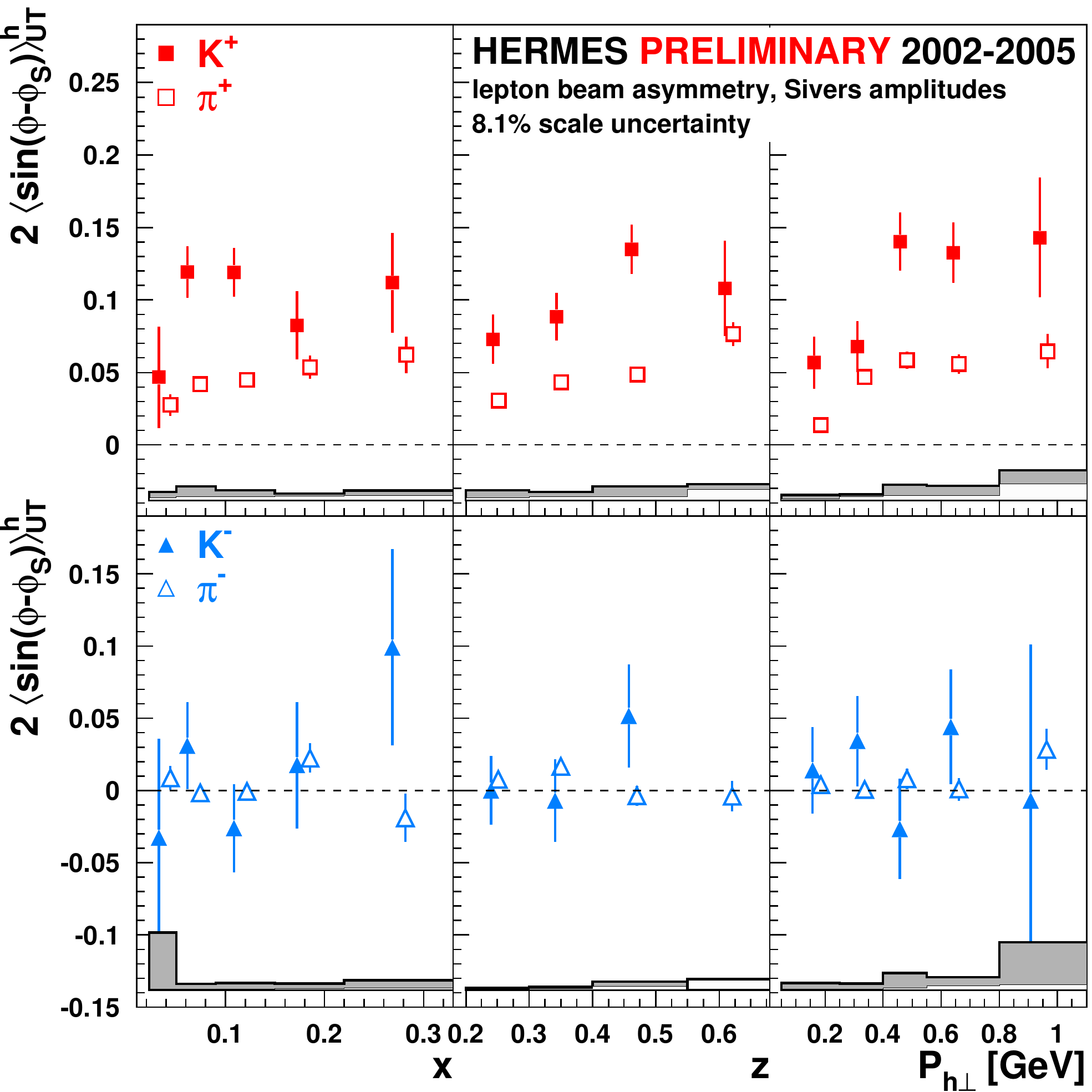}
\caption{Collins (left) and Sivers (right) amplitudes for charged kaons (closed symbols) in comparison to the pion data (open symbols) as a function of $x$, $z$ and $P_{h\perp}$. The error bands give the maximal systematic error; in addition there is a common 8.1\% scaling uncertainty due to the target polarization uncertainty.} \label{Fig:PiK}
\end{figure*}

Figure \ref{Fig:Pions} shows the Collins (left) and Sivers (right) amplitudes for pions.
The Collins amplitude is positive for $\pi^+$, negative for $\pi^-$ and consistent with zero for $\pi^0$.
The magnitude of the $\pi^-$ results appears to be similar or even larger than that of the $\pi^+$ results, which leads to the conclusion that the disfavored Collins fragmentation function has a substantial magnitude with opposite sign to the favored one.
In combination with further information from the unpolarized process $e^+e^- \rightarrow h_1h_2X$ studied at Belle \cite{Belle} the transversity distribution function and the Collins fragmentation function for $u$ and $d$ quarks can be extracted \cite{Anselmino}.
For charged kaons (Fig.~\ref{Fig:PiK} (left)) no significantly non-zero Collins amplitudes were found. 
However, the kaon results are consistent with the pion results within statistical uncertainty.

The Sivers amplitudes are significantly positive for $\pi^+$, (Fig.~\ref{Fig:Pions} right) and $K^+$ (Fig.~\ref{Fig:PiK} right), and consistent with zero for $\pi^-$ as well as $K^-$. 
The positive signal for $\pi^+$ and $K^+$ implies non-vanishing orbital angular momentum of the quarks inside the nucleon.
For both the Collins and the Sivers case, the $\pi$-meson results fulfill isospin symmetry.

\section{STRANGENESS IN THE NUCLEON}

HERMES has extracted the strange quark distribution functions $s(x)+\bar{s}(x)$ and $\Delta (s(x) + \bar{s}(x))$ in leading order $\alpha_s$ using data on DIS off longitudinally polarized deuterium \cite{deltas}, an isoscalar target.
The analysis is based on the assumption of charge conjugation invariance in fragmentation and isospin symmetry between proton and neutron.
Since strange quarks carry no isospin, the strange PDFs can be assumed to be identical for both nucleons.
The fragmentation process in turn can be described by fragmentation functions which have no isospin dependence.
The multiplicity of charged kaons $K=(K^+ + K^-)$ in semi-inclusive DIS from a deuterium target can thus be expressed as
\begin{equation}
	\frac{d N^K(x)}{dN^{\mathrm{DIS}}(x)} = \frac{
	Q(x) \int D_Q^K(z)dz + S(x) \int D_S^K(z)dz
	}{
	5Q(x) + 2S(x)
	},
	\label{Eq:dsmult}
\end{equation}
where $Q(x)\equiv u(x)+\bar{u}(x)+d(x)+\bar{d}(x)$ and $S(x)\equiv s(x) + \bar{s}(x)$ are the strange and non-strange parton distributions.
Furthermore, $D_Q^K(z) \equiv 4D_u^K(z) + D_d^K(z)$ and $D_S^K(z) \equiv 2D_s^K(z)$ are the strange and non-strange fragmentation functions, describing for $D_q^K(z)$ the number density of charged kaons produced from a struck quark of flavor $q$. 

\begin{figure*}[t]
\centering
\includegraphics[width=0.4\textwidth]{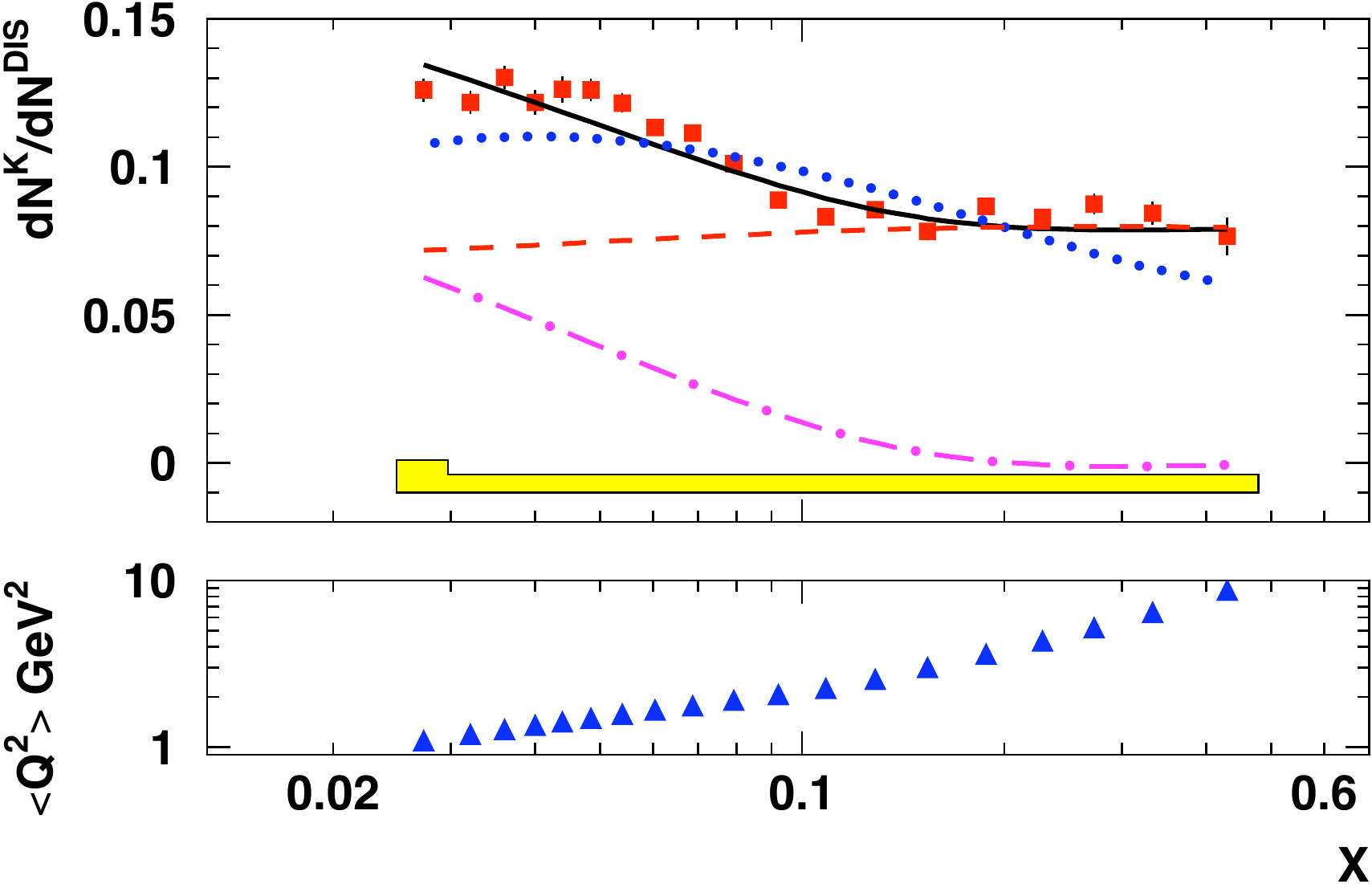}\vspace{0.5cm}
\includegraphics[width=0.49\textwidth]{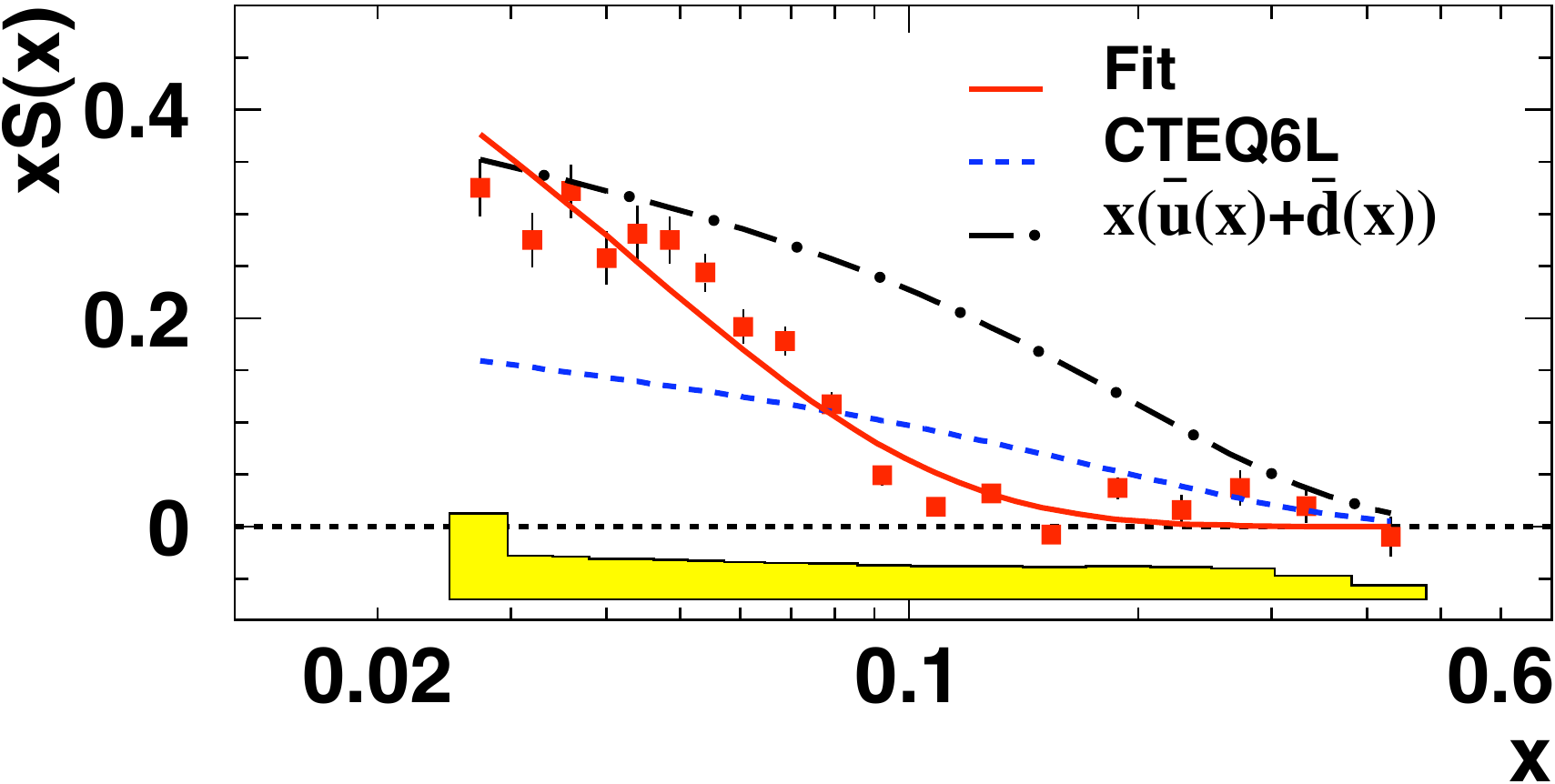}
\caption{Left: Multiplicity of charged kaons in semi-inclusive DIS from a deuterium target as a function of Bjorken x, corrected to $4\pi$. Right: The strange parton distribution function $xS(x)$ extracted from the multiplicities on the left, evolved to $Q_0^2=2.5\,\mathrm{GeV}^2$.} \label{Fig:deltas1}
\end{figure*}

Fig.~\ref{Fig:deltas1} (left) shows this multiplicity as measured at HERMES, after correction for QED radiation, instrumental resolution and acceptance effects using a technique that unfolds kinematic migration of events \cite{g1}.
The kinematic requirements were similar to those of the transversity analysis presented in Sec. \ref{Sec:Trans}, with the exception that here the fractional virtual photon energy $y$ was required to be below $0.85$.
Furthermore, the $z$ range was $0.2-0.8$ and in addition $x_F \approx 2p_L/W > 0.1$, where $p_L$ is the longitudinal momentum of the hadron relative to the virtual photon direction in the photon-nucleon center-of-mass frame.

Using strange and non-strange PDFs from CTEQ6L \cite{CTEQ} and taking the integrated fragmentation functions as free parameters does not yield a fit describing the data (dotted line).
Instead, $S(x)$ was considered unknown, and Eq.~(\ref{Eq:dsmult}) was used to extract the product $S(x)\int D_S^K(z)dz$ (where here and in the following all $z$ integration boundaries are $0.2-0.8$ as given by the cut).
For this, the value of $Q(x)$ was taken from CTEQ6L, while the term $2S(x)$ in the denominator was assumed to be zero in a first iteration (resulting in a small correction in a second step).
Assuming that $S(x)=0$ for $x>0.15$ (where the multiplicity and thus $S(x)/Q(x)$ is approximately constant) yields $\int D_Q^K(z)dz = 0.398\pm0.010$, which is in excellent agreement with the value obtained with the most recent global analysis of fragmentation functions \cite{Sassot}, $0.435\pm0.044$ for $Q^2=2.5\,\mathrm{GeV}^2$.
The value $0.398$ was then used to obtain the product $S(x)\int D_S^K(z)dz$ from Eq.~(\ref{Eq:dsmult}).
The solid line in Fig.~\ref{Fig:deltas1} (left) shows the multiplicity calculated from Eq.~(\ref{Eq:dsmult}) using a fit to $S(x)\int D_S^K(z)dz$
for the strange contribution.
The dashed (dash-dotted) curve singles out the non-strange (strange) contribution.
By applying evolution factors from CTEQ6L the result was evolved to $Q^2=2.5\,\mathrm{GeV}^2$.
Dividing by $\int D_S^K (z) dz=1.27 \pm 0.13$ (calculated from \cite{Sassot})
then yields the strange parton distribution function shown in 
Fig.~\ref{Fig:deltas1} (right). 
The dashed (dash-dotted) curves represent parameterizations of $xS(x)$ and $x(\bar{u}(x)+\bar{d}(x))$ from CTEQ6L.
Independent of the normalization given by \cite{Sassot}, the shape of the HERMES data is incompatible with the shape given by CTEQ6L.

In order to extract the strange quark helicity distribution $\Delta S(x) = \Delta s(x) + \Delta \bar{s}(x)$ the inclusive ($A_{||}$) and semi-inclusive ($A_{||}^{h}$) double-spin asymmetries are used:
\begin{equation}
	A_{||}^{(h)}=
	\frac{
	\sigma^{\stackrel{\rightarrow}{\Leftarrow} , (h)}
	-
	\sigma^{\stackrel{\rightarrow}{\Rightarrow} , (h)}
	}
	{
	\sigma^{\stackrel{\rightarrow}{\Leftarrow} , (h)}
	+
	\sigma^{\stackrel{\rightarrow}{\Rightarrow} , (h)}
	},
\end{equation}
where $\stackrel{\rightarrow}{\Rightarrow}$ ($\stackrel{\rightarrow}{\Leftarrow}$) denotes that beam helicity and target spin are aligned parallel (anti-parallel).
Both asymmetries can be expressed in terms of the non-strange quark helicity distribution $\Delta Q(x)=\Delta u(x) + \Delta \bar{u}(x) + \Delta d(x) + \Delta \bar{d}(x)$ and the strange quark helicity distribution $\Delta S(x) = \Delta s(x) + \Delta \bar{s}(x)$ as
\begin{eqnarray}
	A_{{||},d}(x)\frac{d^2 N^{\mathrm{DIS}}(x)}{dxdQ^2}
	&=&
	\mathcal{K}_{LL}(x,Q^2)\left[ 5 {\Delta Q(x)} + 2{\Delta S(x)} \right]\label{Eq:Ainc}\\
	A_{||,d}^{\mathrm{K}}(x)
	\frac{d^2 N^{\mathrm{K}}(x)}{dxdQ^2}
	&=&
	\mathcal{K}_{LL}(x,Q^2)\left[{\Delta Q(x)} \int D_Q^{\mathrm{K}}(z)dz + {\Delta S(x)}\int D_S^{\mathrm{K}}(z) dz \right],
		\label{Eq:Asinc}
\end{eqnarray}
where $\mathcal{K}_{LL}$ is a kinematic factor containing the hard scattering cross section.
Eqs.~(\ref{Eq:Ainc}) and (\ref{Eq:Asinc}) allow the simultaneous extraction of the helicity distributions $\Delta Q(x)$ and $\Delta S(x)$, making use of the non-strange integrated fragmentation function extracted from the multiplicities obtained from the same data.

\begin{figure*}[b]
\centering
\includegraphics[width=0.49\textwidth]{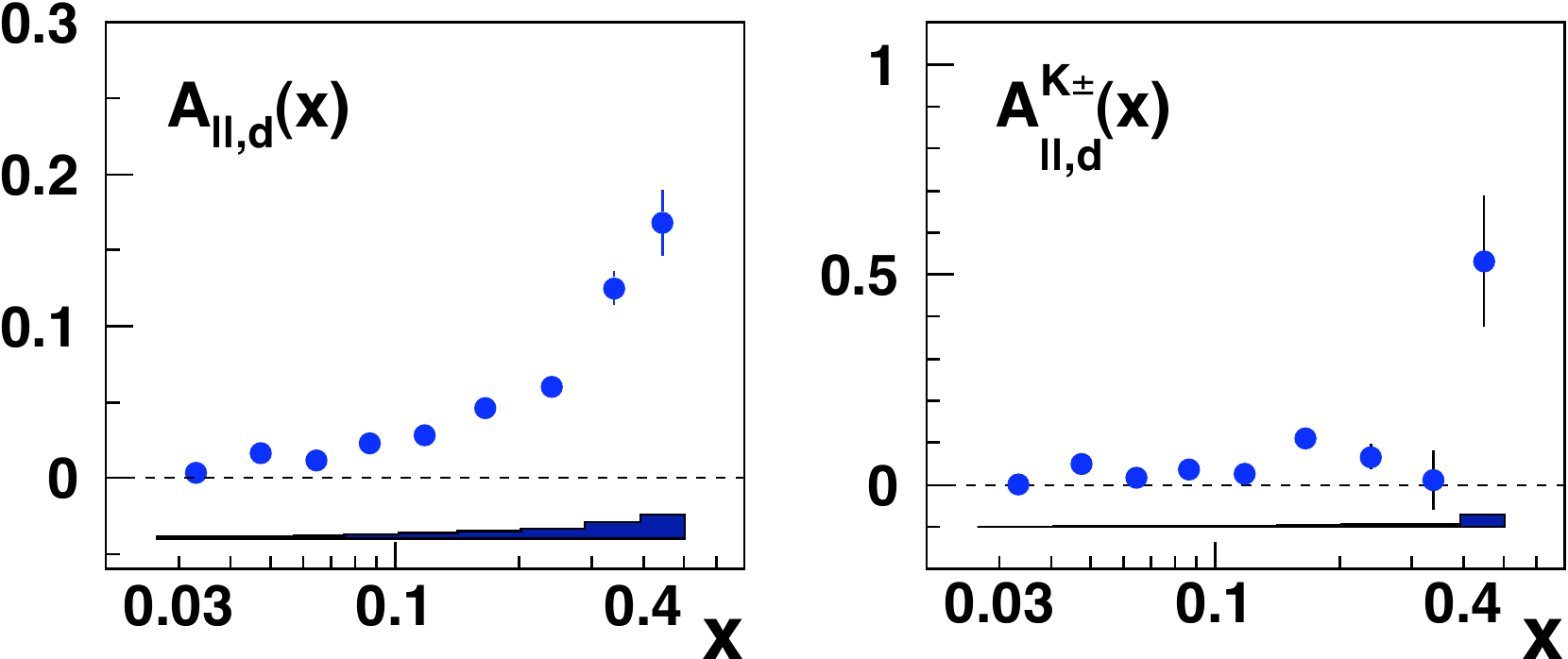}\hspace{1cm}
\includegraphics[width=0.35\textwidth]{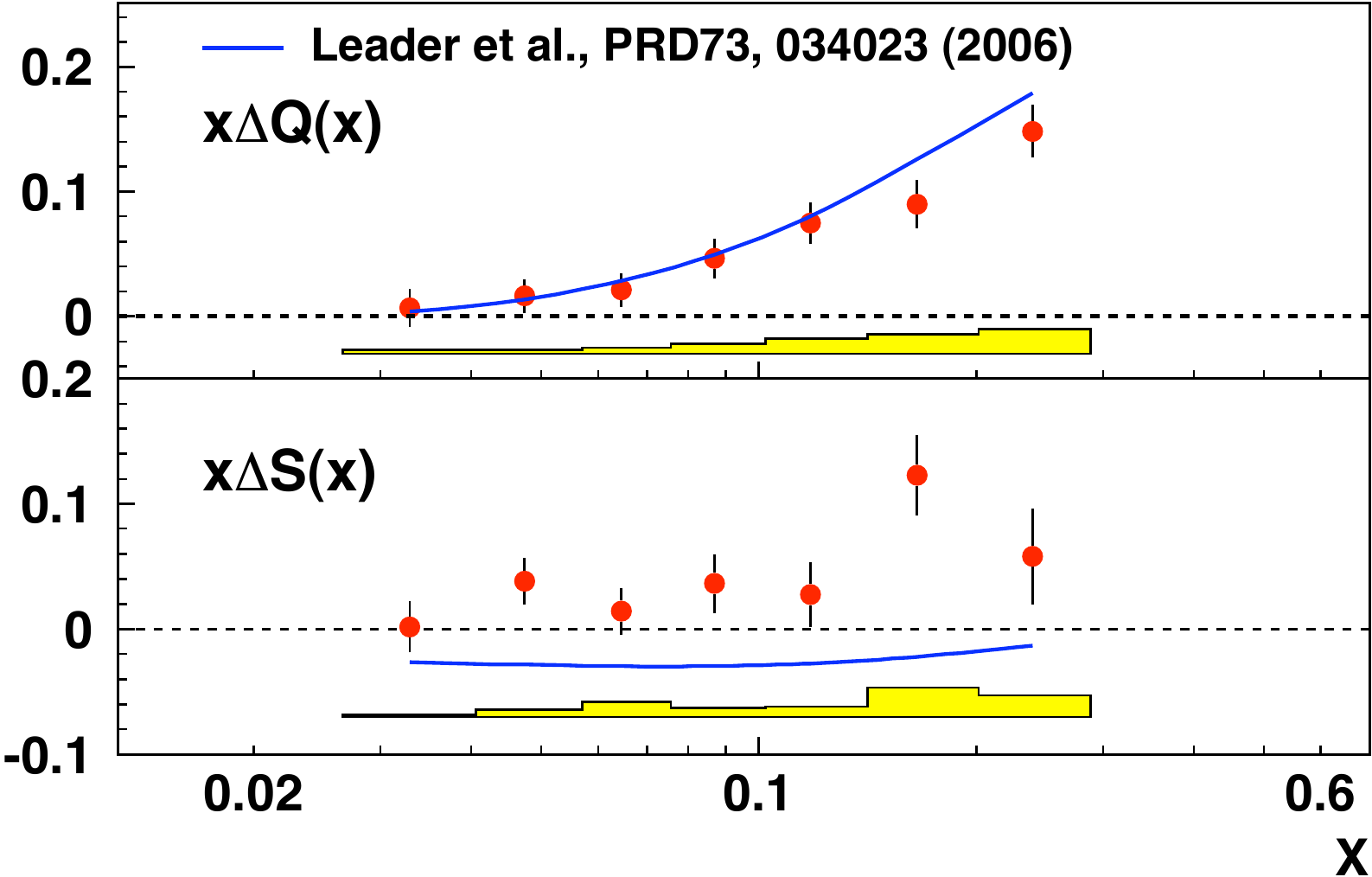}
\caption{Left panels: Lepton-nucleon polarized cross section asymmetries for inclusive DIS ($A_{||,d}$) and charged kaons in semi-inclusive DIS ($A_{||,d}^K$) from a deuteron target. Right panel: Non-strange and strange quark helicity distributions at $Q_0^2=2.5\,\mathrm{GeV}^2$ as a function of Bjorken $x$. The curves are the LO results of Leader et al. from their analysis of world data. Error bars show the statistical, error bands the systematic uncertainties.} \label{Fig:deltas2}
\end{figure*}

The asymmetries are shown in Fig.~\ref{Fig:deltas2} (left). They were corrected for QED radiation and instrumental smearing in the same way as the multiplicities. 
The systematic error includes contributions from the beam and target polarizations, the neglect of the transverse spin structure function $g_2(x)\approx 0$ \cite{g2} and, in the case of $A_{||,d}^K$, from kaon misidentification. 

Fig.~\ref{Fig:deltas2} (right) shows the helicity distributions obtained from 
the asymmetries using Eqs.~(\ref{Eq:Ainc}) and (\ref{Eq:Asinc}).
As for $S(x)$, the value of $\int D_S^K(z)dz=1.27 \pm 0.13$ was used to extract $\Delta S(x)$. The result agrees well with the less precise data in \cite{hermes_dq}, and is consistent with zero over the measured range.

The first moment of $\Delta Q$ obtained by integration over the measured range of $0.02<x<0.6$ is $0.359\pm 0.026\mathrm{(stat.)}\pm0.018\mathrm{(sys.)}$, in good agreement with the full moment $0.381\pm 0.010\mathrm{(stat.)} \pm 0.027\mathrm{(sys.)}$ extracted from HERMES $g_{1,d}$ data \cite{g1}. 
The corresponding first moment of $\Delta S$ is  $0.037\pm0.019\mathrm{(stat.)}\pm0.027\mathrm{(sys.)}$, in not serious disagreement with the inclusive HERMES measurement ($\Delta S=-0.0435\pm0.010\mathrm{(stat.)}\pm0.004\mathrm{(sys.)}$).
This yields for the octet combination $\Delta q_8= \Delta Q - 2\Delta S=0.285\pm0.046\mathrm{(stat.)}\pm0.057\mathrm{(syst.)}$, which is substantially less than the value of the axial charge $a_8\equiv\Delta q_8 = \int_0^1 \Delta q_8(x)dx=0.586\pm0.031$ extracted from the hyperon decay constants assuming SU(3) symmetry \cite{hyperon}.
Possible reasons for the low value of $\Delta q_8$ could be either a missing contribution from low $x$ below the measured range or a violation of SU(3) symmetry.


\end{document}